\documentclass[twocolumn,amsmath,amssymb]{jpsj3}
\usepackage{txfonts}
\usepackage{graphicx}

\title{Andreev Reflection in Weyl Semimetals}

\author{
Shuhei Uchida$^1$, Tetsuro Habe$^1$, and Yasuhiro Asano$^{1,2}$}
\inst{$^1$Department of Applied Physics, Hokkaido University, Sapporo 060-8628, Japan \\
$^2$Center for Topological Science \& Technology, Hokkaido University, Sapporo 060-8628, Japan} %\\

\abst{We theoretically study low energy electric transport in a junction consisting of a Weyl semimetal 
and a metallic superconductor. The characteristic features of the differential conductance depend on the 
relative directions between the current and the vector connecting the two Weyl points.
When the electric current is perpendicular to the vector, the conductance spectra are sensitive to 
the direction and the amplitude of magnetic moment at the junction interface. 
This is a direct consequence of the chiral spin configuration on the fermi surface
near the Weyl points.
}

\begin{document}
\maketitle

\section{Introduction}
Weyl semimetals are a novel topological material in 
three-dimension~\cite{murakami,wan,burkov1,burkov,halaz,ojanen,zyuzin1}. 
In the bulk, Weyl semimetal has the gapless spectra characterized by even number 
of Weyl points with opposite topological charges in Brillouin zone, which 
leads to unusual transport 
properties~\cite{chen,wei,yang,zyuzin2,hosur,ashby,rosenstein,ominato}.
Topological properties in these cases result
from the separation of the individual Weyl points in the absence of
 of either time-reversal or inversion symmetry.
Various theoretical model has been discussed to realize the semimetal 
phase~\cite{jiang,cho,delplace}.
In experiments, the semimetal is realized in multilayers of GeTe/Sb$_2$Te$_3$~\cite{simpson,sa}
and Cd$_3$As$_2$~\cite{neupane,borisenko,wang}.
Recently a possibility of superconductivity has been discussed in several 
theoretical studies~\cite{meng,wei2,cho2}. Unconventional Cooper pairing symmetry would be expected 
because of the chiral spin structure on the fermi surface. 
Although finding the superconductivity within existing semimetals might be 
difficult, realizing the superconducting 
correlation there is basically possible by injecting Cooper pairs
due to the proximity effect. As the first step in this research direction, we address 
the Andreev reflection in Weyl semimetal in the present paper.
%Experimental observation of such anomalous properties has been desired.

In this paper, we theoretically study the low energy transport through a junction consisting a Weyl semimetal 
and a metallic superconductor. The differential conductance of the Wely-semimetal/superconductor
junction is calculated from the normal and the Andreev reflection coefficients of the junction. 
We consider a $2\times 2$ simple Hamiltonian which describes the electronic structure of a Weyl semimetal 
breaking the time-reversal symmetry. Within this model, the semimetallic excitation is characterized by 
the linear energy-momentum dispersion relations near the two separated Weyl points. 
The spin structures at the two fermi surfaces are characterized by the opposite 
spin chiral texture to each other. 
When the two Weyl points stay at $\pm \boldsymbol{K}_0$ with $\boldsymbol{K}_0 =(0,0, k_0) $,
the topological surface states appears on the four surfaces parallel 
to $\boldsymbol{K}_0$ direction.
We consider two types of junction: the current parallel to $\boldsymbol{K}_0$ and the current 
perpendicular to $\boldsymbol{K}_0$. When the potential barrier at the interface 
is spin-independent, the conductance spectra in the junctions are similar to those 
in the usual normal-metal/superconductor junctions. When 
the potential at the interface is spin active, on the other hand,  
the conductance spectra depend sensitively on types of the junction and directions of the magnetic moment.  
In particular, for the current perpendicular to $\boldsymbol{K}_0$, both the normal and the Andreev reflections 
are suppressed due to the chiral spin structure on the fermi surface. The spin-flip potentials at the interface 
relax the spin mismatch in the reflection process.

\section{Weyl Semimetal}
\begin{figure}
\includegraphics[width=8cm]{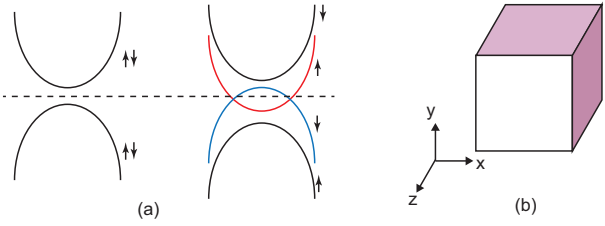}
\caption{(a) Schematic band structures of theoretical model. The Zeeman effect shifts the bands depending 
on their spin. We consider two bands on the fermi level indicated by a broken line.
(b) We consider two types of junction: current parallel to the $z$ axis and that parallel to the $x$.
The topological surface states appear on the surfaces perpendicular to the $x$ axis and those to the $y$ axis.
}
\label{fig1}
\end{figure}

To describe the electronic states in Weyl semimetals, we use a simple model given by~\cite{yang}
\begin{align}
H_{W} =& \sum_{\alpha, \beta} \int d\boldsymbol{r} \; 
\psi_\alpha^\dagger(\boldsymbol{r}) \left[
-\frac{\hbar^2}{2m_W}(\nabla^2+k_0^2) \hat{\sigma}_z \right. \nonumber\\
&\left. -i \lambda( \partial_x\hat{\sigma}_x + \partial_y\hat{\sigma}_y)
-\mu_W \hat{\sigma}_0 \right]_{\alpha,\beta} 
\psi_\beta(\boldsymbol{r}), \label{hw}
\end{align}
where $\psi^\dagger_\alpha(\boldsymbol{r})$ ($\psi_\alpha(\boldsymbol{r}) )$ is the creation (annihilation) 
operator of an electron with spin $\alpha$ at $\boldsymbol{r}$, $\nabla$ is the three-dimensional Laplacian, 
$m_W$ is the effective mass of an electron, $\lambda$ denotes 
the coupling constant of the spin-orbit interaction, and $\mu_W$ is the chemical potential 
measured from the Weyl point.
We originally begin with the spin-degenerate two-band model as shown in Fig.~\ref{fig1}(a). 
The Zeeman field decreases (increases) the energy of the spin-up (spin-down) band.
Large enough Zeeman fields result in
the inverted band structure. The effects of the Zeeman field 
is taken into account through $k_0$ in Eq.~(\ref{hw}). The Pauli's matrices $\hat{\sigma}_j$ for $j=x, y$ 
and $z$ represent the real spins of an electron. The unit matrix in spin space is $\hat{\sigma}_0$.
By neglecting the two bands away from the 
fermi level, the electric structure are described by Eq.~(\ref{hw}).
In the Fourier representation, Eq.~(\ref{hw}) becomes
\begin{align}
H_W(\boldsymbol{k}) = \left[
\begin{array}{cc}
\epsilon_{\boldsymbol{k}}- \mu_W & \lambda(k_x- i k_y) \\
\lambda(k_x + i k_y) & -\epsilon_{\boldsymbol{k}}- \mu_W
\end{array}\right],
\end{align}
with
$ \epsilon_{\boldsymbol{k}} = (\hbar^2/2m_W)( \boldsymbol{k}^2-k_0^2)$. 
The energy dispersion and the wave functions are obtained as 
\begin{align}
\left(\begin{array}{c} \alpha_{\boldsymbol{k}} \\ \beta_{\boldsymbol{k}} (k_x+ik_y)/p
\end{array}\right), \quad  \left(\begin{array}{c} -\beta_{\boldsymbol{k}}(k_x-ik_y)/p \\ \alpha_{\boldsymbol{k}} 
\end{array}\right),\label{wplus}
\end{align}
for $E^W_{\boldsymbol{k}} - \mu_W$ and $-E^W_{\boldsymbol{k}} - \mu_W$, respectively.
Here we define following quantities,
\begin{align}
E^W_{\boldsymbol{k}}=& \sqrt{\epsilon^2_{\boldsymbol{k}} + (\lambda p)^2}, \quad \boldsymbol{p}=(k_x, k_y,0),  \\
\alpha_{\boldsymbol{k}}=& \sqrt{ \frac{1}{2}\left( 1+\frac{\epsilon_{\boldsymbol{k}}}{E^W_{\boldsymbol{k}}} \right)},
\quad
\beta_{\boldsymbol{k}}=\sqrt{ \frac{1}{2}\left( 1-\frac{\epsilon_{\boldsymbol{k}}}{E^W_{\boldsymbol{k}}} \right)}.
\end{align}
The two Weyl poins appear at $\pm \boldsymbol{K}_0$.
When we consider low energy transport around the chemical potential, 
the energy band with $E^W_{\boldsymbol{k}} - \mu_W$ 
carries the electric current. The group velocity of Weyl semimetal is anisotropic.
The velocities are represented by
\begin{align}
v_{x(y)}= \frac{\hbar k_{x(y)}}{m_W} \frac{\epsilon_{\boldsymbol{k}}}{E^W_{\boldsymbol{k}}} 
+ \frac{\lambda^2 k_{x(y)}}{ \hbar E^W_{\boldsymbol{k}} },\quad
v_z= \frac{\hbar k_z}{m_W} \frac{ \epsilon_{\boldsymbol{k}} } { E^W_{\boldsymbol{k}} },
\end{align}
in the $x(y)$ and the $z$ direction, respectively. In addition, the expectation value of 
spin $\boldsymbol{S} = (\hbar/2) \hat{\boldsymbol{\sigma}}$ are calculated to be
\begin{align}
\left\langle S_{x(y)} \right\rangle = \frac{\lambda k_{x(y)}}{E^W_{\boldsymbol{k}}},
\quad 
\left\langle S_{z} \right\rangle = \frac{\epsilon_{\boldsymbol{k}}}{E^W_{\boldsymbol{k}}},
\label{spin}
\end{align} 
in units of $\hbar/2$.

The Hamiltonian in the hole space is represented by $-H_W^\ast(-\boldsymbol{k})$.
The wave function at $E=-E_{\boldsymbol{k}}+\mu_W$ is given by
\begin{align}
\left( 
\begin{array}{c} \alpha_{\boldsymbol{k}} \\
- \beta_{\boldsymbol{k}} (k_x-ik_y)/p
\end{array}
\right).
\end{align}
The spin expectation values are calculated to be
\begin{align}
\left\langle S_{x(y)} \right\rangle = \frac{-\lambda k_{x(y)}}{E^W_{\boldsymbol{k}}},
\quad 
\left\langle S_{z} \right\rangle = \frac{\epsilon_{\boldsymbol{k}}}{E^W_{\boldsymbol{k}}},
\label{spinh}
\end{align} 
in the hole space. 

\subsection{Fermi surface}
\begin{figure}
\includegraphics[width=8cm]{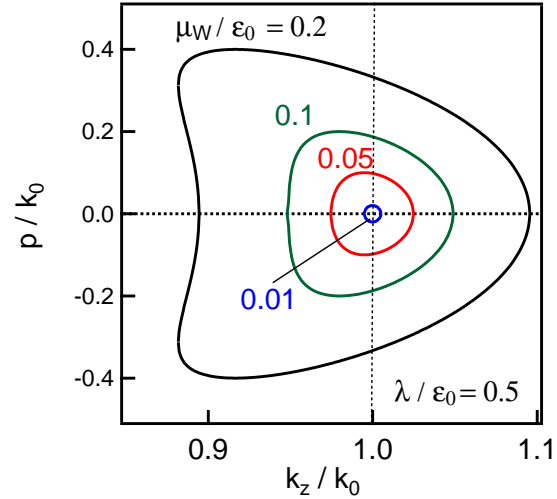}
\caption{ The fermi surface around $\boldsymbol{K}_0$ are plotted at $\lambda/\epsilon_0=0.5$ for several $\mu_W$.
The shape of the fermi surface is independent of the direction of momenta 
$\boldsymbol{p}=(k_x,k_y,0)$ in the $xy$ plane.
%When we choose $\mu_W<0.4 \epsilon_0$, fermi surface encloses the Weyl points at $(0,0,k_0)$.
%The shape of the fermi surface becomes more ellipsoidal for smaller $\mu_W$. 
}
\label{fig2}
\end{figure}

To study transport properties unique to the Weyl semimetals, we need to set the chemical potential
$\mu_W$ to be small enough values. 
We show the shape of the fermi surface at $\lambda=0.5\epsilon_0$ 
for several $\mu_W$ in Fig.~\ref{fig2}, where the line connects the equal 
energy points in the Brillouin zone.
For $\mu_W/\epsilon_0 < 0.4$, the two disconnected fermi surfaces 
 enclose the two Weyl points at $(0,0,\pm k_0)$. 
We only show the fermi surface around $\boldsymbol{K}_0=(0,0,k_0)$ in Fig.~\ref{fig2}.
At $\mu_W/\epsilon_0 = 0.2 $, the shape of fermi surface is still distorted.
The fermi surface becomes more ellipsoidal for smaller $\mu_W$. 
In this paper, we fix the parameters as  $\lambda=0.5\epsilon_0$ and  $\mu_W=0.1\epsilon_0$.

\subsection{Wave function at a fixed energy}
It is possible to consider two types of junction: (i) the current in the $z$ direction 
and (ii) the current in the $x$ direction. We first discuss the wave function for the current
in the $z$ direction.
At an energy $E >0$, such wave function in the $z$ direction proportional to 
$e^{i(k_x x+ k_y y)}$ is described by
\begin{align}
&\Psi_W(z\leq 0)
=
\frac{1}{z_+^e}\left[\begin{array}{c} f_e^z \\ \gamma_+ \\ 0 \\ 0 \end{array}\right] 
\left( e^{ik_e^+z} a^+ + e^{-ik_e^+z} A^+\right)\nonumber\\
&+
\frac{1}{z_-^e}\left[\begin{array}{c} \gamma_- \\ f_e^z \\ 0 \\ 0 \end{array}\right] 
\left( e^{-ik_e^-z} a^-+e^{ik_e^-z} A^-\right)\nonumber\\
&+
\frac{1}{z_+^h}\left[\begin{array}{c} 0\\ 0\\ f_h^z \\ -\gamma_- \end{array}\right] 
\left( e^{-ik_h^+z} b^+ + e^{ik_h^+z} B^+\right) \nonumber\\
&+
\frac{1}{z_-^h}\left[\begin{array}{c} 0\\ 0\\  -\gamma_+\\ f_h^z \end{array}\right] 
\left( e^{ik_h^-z} b^-+e^{-ik_h^-z} B^-\right), \label{wf_z}
\end{align}
\begin{align}
f_{e(h)}^z=&\mu_W+(-)E+\sqrt{D^z_{e(h)}}, 
%\; f_h^z=\mu_W-E+\sqrt{D_h}, 
\quad \gamma_\pm=\lambda(k_x\pm ik_y), \\
k_{e(h)}^\pm=& \sqrt{ k_0^2 - p^2 \pm (2m_W/\hbar^2) \sqrt{D^z_{e(h)}}}, \\
%\; k_h^\pm= \sqrt{ k_0^2 - p^2 \pm \sqrt{D_h}}, \\
D^z_{e(h)}=& \left\{ \mu_W + (-) E \right\}^2 - (\lambda p)^2,
%\; D_h= (\mu_W-E)^2 - (\lambda p)^2,
\end{align}
where $z_\pm^e$ and $z_\pm^h$ are factors which normalize the wave functions.
The coefficients $a^\pm$ ($b^\pm$) are the amplitudes of wave function 
incoming into the junction interface as an electron (hole). 
While 
$A^\pm$ ($B^\pm$) are the amplitudes of wave function 
outgoing from the junction interface as an electron (hole). 
We note that $E^W_{\boldsymbol{k}}$ was replaced by $\mu_W+E$ in the electron space.
At the same time, $\epsilon_{\boldsymbol{k}}$ was replaced by $\pm D^z_e$ for $k^\pm_e$ channel.
In the hole space, we have applied relations $E^W_{\boldsymbol{k}} \to \mu_W-E$ and 
$\epsilon_{\boldsymbol{k}} \to \pm D^z_h$ for $k^\pm_h$ channel. 
In this way, we obtain Eq.~(\ref{wf_z}).

When the current flows in the $x$ direction, on the other hand, the wave function in the 
$x$ direction proportional to $e^{i(k_yy + k_z z)}$ becomes
\begin{align}
&\Psi_W(x\leq 0)
=
\frac{1}{z_{k^+_e}}\left[\begin{array}{c} f_{e,+}^x \\ \gamma_{k^+_e,+} \\ 0 \\ 0 \end{array}\right] 
 e^{ik_e^+x} a^+ 
+
\frac{1}{z_{-k^-_e}}\left[\begin{array}{c} \gamma_{-k^-_e,-} \\ f_{e,-}^x \\ 0 \\ 0 \end{array}\right] 
e^{-ik_e^-x} a^- \nonumber\\
&+
\frac{1}{z_{-k^+_e}} \left[\begin{array}{c} f_{e,+}^x \\ \gamma_{-k^+_e,+} \\ 0 \\ 0 \end{array}\right] 
e^{-ik_e^+x} A^+ 
+
\frac{1}{z_{k^-_e}} \left[\begin{array}{c} \gamma_{k^-_e,-} \\ f_{e,-}^x \\ 0 \\ 0 \end{array}\right] 
e^{ik_e^-x} A^- \nonumber\\
&+
\frac{1}{ z_{-k^+_h} }
\left[\begin{array}{c} 0\\ 0\\ f_{h,+} \\ -\gamma_{-k^+_h, -} \end{array}\right] 
e^{-ik_h^+x} b^+ 
+
\frac{1}{z_{k^-_h}} \left[\begin{array}{c} 0\\ 0\\  -\gamma_{k_h^-,+}\\ f_{h,-} \end{array}\right] 
e^{ik_h^-x} b^- \nonumber\\
&+
\frac{1}{ z_{k^+_h} }
\left[\begin{array}{c} 0\\ 0\\ f_{h,+} \\ -\gamma_{k^+_h,-} \end{array}\right] 
 e^{ik_h^+x} B^+
+
\frac{1}{z_{-k^-_h}} \left[\begin{array}{c} 0\\ 0\\  -\gamma_{-k_h^-,+}\\ f_{h,-} \end{array}\right] 
e^{-ik_h^-x} B^-, \label{psiz}
\end{align}
\begin{align}
f_{e(h),\pm}^x=& \mu_W+(-) E+ \sqrt{ D^x_{e(h)} } \mp \lambda k_0 \tilde{\lambda}^2/2,\\
k_{e(h)}^\pm=& \sqrt{ A_0 \pm (2m_W/\hbar^2) \sqrt{D^x_{e(h)} }}, \\ 
%\; f_{h,\pm}^z=\mu_W-E+ \sqrt{D_h} \mp (\lambda k_0)^2/(2\epsilon_0),\\
\gamma_{k,\pm}=&\lambda(k \pm ik_y),\;\;  \tilde{\lambda}=\lambda k_0/\epsilon_0, \;\; \epsilon_0 = \hbar^2k_0^2/(2m_W),\\
%\; k_h^\pm= \sqrt{ A_0 \pm 2m_W\sqrt{D_h}/\hbar^2}, \\
A_0=&k_0^2-k_y^2-k_z^2 - k_0^2 \tilde{\lambda}^2/2,\\
D^x_{e(h)}=& \lambda^2(k_z^2-k_0^2) + (\lambda k_0)^2 \tilde{\lambda}^2/4 + \left\{ \mu_W+(-) E \right\}^2.
%&D_h= \lambda^2(k_z^2-k_0^2) + (\lambda k_0)^4/(4\epsilon_0^2) + (\mu_W-E)^2,
\end{align}
On the way to Eq.~(\ref{psiz}), we have used relations $E_{\boldsymbol{k}} \to \mu_W + E$ and
 $\epsilon_{\boldsymbol{k}} \to -\tilde{\lambda} \lambda k_0/2 \pm \sqrt{D^x_e}$ for $k^\pm_e$ channel
in the electron space. 
In the hole space, we have applied $E_{\boldsymbol{k}} \to \mu_W - E$ and
 $\epsilon_{\boldsymbol{k}} \to -\tilde{\lambda} \lambda k_0/2 \pm \sqrt{D^x_h}$ for $k^\pm_h$ channel.

\subsection{Surface bound states}
The topologically protected surface bound states appear at the surfaces 
perpendicular to the $x$ axis and those perpendicular to the $y$ axis. 
However it is absent on the surface perpendicular to the $z$ axis.
It is possible to study the spectra of such topological surface states by using the wave function in Eq.~(\ref{psiz}).
In the electron space, the wave 
function of the bound states neat the surface (i.e., $ x= 0$) can be described 
by  
\begin{align}
&\Psi_W^e(x)
=
\left[\begin{array}{c} f_{e,+}^x \\ \gamma_{-k^+_e,+} \end{array}\right] 
e^{-ik_e^+x} A^+ 
+
\left[\begin{array}{c} \gamma_{k^-_e,-} \\ f_{e,-}^x  \end{array}\right] 
e^{ik_e^-x} A^-.
\end{align}
These two wave functions represent the wave decaying into the Weyl semimetal ($x<0$) for $D^x_e<0$.
The condition $D^x_e<0$ results in 
the complex wave number in the $x$ direction. 
By imposing the boundary condition $\Psi_W^e(0)=0$,
we obtain the dispersion of the surface bound states as 
$E_{\text{BS}}=\lambda k_y - \mu_W$ for $k^2_y < k_0^2(1 - \tilde{\lambda}^2/4) - k_z^2$. 
Thus chiral electric current flows in the $y$ direction. 
The wave function of the bound states are obtained as
\begin{align}
\Psi_{BS}(x) = C_0 e^{\tilde{\lambda} k_0 x /2} \sin\left( \sqrt{k_0^2-k_z^2-k_y^2}x \right)
\left( \begin{array}{c} \delta_0 \\ \delta_0^\ast \end{array} \right),
\end{align}
for $x<0$, where $\delta_0=e^{i\pi/4}$ and $C_0$ is a constant.
We have used a relation $\tilde{\lambda}\ll 1$.
The bound states are the eigen states of $\hat{\sigma}_y$.
At $k_y= \mu/\lambda$, a bound state appears on the fermi level.
Due to the complex wave number under $D^x_e<0$, unfortunately, 
the surface bound state does not contribute to the electric transport 
in the $x$ direction.
In the ballistic regime, the wave numbers in the parallel 
directions to the interface are conserved in the transmission 
and the reflection processes. 
An electron incoming into the interface through a propagation channel 
cannot be reflected or transmitted via the bound state.
This is because the bound states is formed by the decaying waves belonging to 
the evanescent channel.

\subsection{Junction with a superconductor }
The Hamiltonian in a metallic superconductor is represented in the momentum space,
\begin{align}
H_S(\boldsymbol{k})=& \left[ 
\begin{array}{cc} 
\xi_{\boldsymbol{k}} \hat{\sigma}_0 & \Delta i \hat{\sigma}_y \\
- \Delta i \hat{\sigma}_y & -\xi^\ast_{\boldsymbol{-k}}\hat{\sigma}_0
\end{array}
\right],\\
\xi_{\boldsymbol{k}}=&\frac{\hbar^2\boldsymbol{k}^2}{2m_S} - \mu_S
\end{align}
where $\Delta$ is the amplitude of the pair potential, and $m_S$ ($\mu_S$) is the mass of an electron (the chemical potential) 
in the superconductor.
The wave function in the $z$ direction, for example, is represented by 
\begin{align}
\Psi_S(z) =& 
\left[ \begin{array}{c} u_0 \hat{\sigma}_0 \\ v_0(-i\hat{\sigma}_y ) \end{array}\right]
\left( \begin{array}{c} C_{\uparrow} \\ C_{\downarrow} \end{array} \right)
e^{i q^e z} \nonumber\\
&+
\left[ \begin{array}{c} v_0(i\hat{\sigma}_y) \\ u_0 \hat{\sigma}_0  \end{array}\right]
\left( \begin{array}{c} D_{\uparrow} \\ D_{\downarrow} \end{array} \right)
e^{-i q^h z},
\end{align}
\begin{align}
u_0 =& \sqrt{ \frac{1}{2}\left(1+\frac{\Omega}{E}\right)},\; \;
v_0 = \sqrt{ \frac{1}{2}\left(1-\frac{\Omega}{E}\right)},\\
\Omega=& \sqrt{E^2-\Delta^2},\;\;
q^{e(h)}= \sqrt{ k_F^2-p^2 +(-) 2m_S\Omega/\hbar^2}.
\end{align}

The wave functions on either sides of the junction are connected by the boundary conditions,
\begin{align}
&\Psi_W(z=0) = \Psi_S(z=0),\label{bc1}\\
&\left.-\frac{\hbar^2}{2m_W}\left[\begin{array}{cc}\hat{\sigma}_z & 0 \\ 0 & \hat{\sigma}_z \end{array}\right]\partial_z 
\Psi_W(z)\right|_{z=0} \nonumber\\
&= \left.- \frac{\hbar^2}{2m_S}\partial_z \Psi_S(z)\right|_{z=0} + 
\left[ V_0 \hat{\sigma}_0 + \boldsymbol{V}\cdot \hat{\boldsymbol{\sigma}} \right]\Psi_S(z=0), \label{bc2}
\end{align}
for the current in the $z$ direction. Here we introduce the barrier potential $V_0\delta(z)$ 
and the magnetic potential $\boldsymbol{V}\cdot \hat{\boldsymbol{\sigma}}\delta(z)$ at the interface. 
For the current in the $x$ direction, we change $z$ to $x$ in the above conditions.
In addition to this, we need to add
\begin{align}
\frac{\lambda}{2i} 
\left[\begin{array}{cc}\hat{\sigma}_x & 0 \\ 0 & -\hat{\sigma}_x \end{array}\right]\Psi_W(x=0)
\end{align}
on the left hand side of Eq.~(\ref{bc2}) to satisfy the current conservation law.
By using the boundary conditions in Eqs.~(\ref{bc1}) and (\ref{bc2}), it is possible to 
obtain the reflection matrix,
\begin{align}
\left[ \begin{array}{c} 
A^+\\A^- \\ B^+ \\ B^- \end{array} \right]
=\left[ \begin{array}{cc} \hat{r}_{ee} & \hat{r}_{eh} \\
\hat{r}_{he} & \hat{r}_{hh}
\end{array} \right]
\left[ \begin{array}{c} 
a^+\\a^- \\ b^+ \\ b^- \end{array} \right],
%\left[ \begin{array}{c} 
%C_\uparrow\\ C_\downarrow \\ D_\uparrow \\ D_\downarrow \end{array} \right]
%=\left[ \begin{array}{cc} \hat{t}_{ee} & \hat{t}_{eh} \\
%\hat{t}_{he} & \hat{t}_{hh}
%\end{array} \right]
%\left[ \begin{array}{c} 
%a^+\\a^- \\ b^+ \\ b^- \end{array} \right].
\end{align}

The differential conductance of the junction is calculated as~\cite{blonder,takane}
\begin{align}
G_{\text{WS}}(eV) =& \left.\frac{e^2}{h} \sum_{\boldsymbol{P}}
\text{Tr}\left[ \hat{1}_{ee} - \hat{R}_{ee}\hat{R}_{ee}^\dagger + \hat{R}_{he}\hat{R}_{he}^\dagger \right]\right|_{E=eV}
,\end{align}
\begin{align}
\hat{R}_{ee} =& \left(\begin{array}{cc} \sqrt{v_e^+} & 0 \\ 0 & \sqrt{v_e^-} \end{array} \right)
\hat{r}_{ee} \left(\begin{array}{cc} 1/\sqrt{v_e^+} & 0 \\ 0 & 1/\sqrt{v_e^-} \end{array} \right),\\  
\hat{R}_{he} =& \left(\begin{array}{cc} \sqrt{v_h^+} & 0 \\ 0 & \sqrt{v_h^-} \end{array} \right)
\hat{r}_{he} \left(\begin{array}{cc} 1/\sqrt{v_e^+} & 0 \\ 0 & 1/\sqrt{v_e^-} \end{array} \right),  \\
\hat{1}_{ee}=&\left(\begin{array}{cc} s_e^+ & 0 \\ 0 & s_e^- \end{array} \right),
\end{align}
where $v_{e}^\pm$ ($v_{h}^\pm$) is the velocity of $k_e^\pm$ ($k_h^\pm$) channel at the electron (hole) 
space in the Weyl semimetal and $\boldsymbol{P}$ denotes the momenta parallel to the interface. 
We define that $s_e^{\pm}$ is 1 for the propagating $k_e^\pm$ channel  
and 0 for the evanescent $k_e^\pm$ channel in the electron space.

\section{Differential Conductance}
Throughout this paper, we fix material parameters in the superconductor as $m_S=m_W$ and $\mu_S=2\epsilon_0$.
These parameters only modify the transmission probability of the junction.
The amplitude of the pair potential is fixed at $\Delta=0.01\epsilon_0$ which gives the smallest 
energy scale in our model.  
As we discussed in Sec.~2.1, we choose $\mu_W=0.1\epsilon_0$ and $\lambda=0.5\epsilon_0$ in the Weyl semimetal.
The barrier potentials are parameterized as $z_0=(m_W/\hbar^2k_0)V_0$ and $\boldsymbol{M}=(m_W/\hbar^2k_0)\boldsymbol{V}$.

We first discuss the differential conductance without magnetic barrier at the interface (i.e., $\boldsymbol{M}=0$).
Fig.~\ref{fig2} shows the differential conductance of Weyl-semimetal/superconductor junction 
for the current parallel to the $z$
axis in (a) and that for the current parallel to the $x$ axis in (b). 
The results are normalized to the conductance of Weyl-semimetal/normal-metal junction 
($G_{\textrm{WN}}$)
at $eV=0$.
\begin{figure}
\includegraphics[width=8cm]{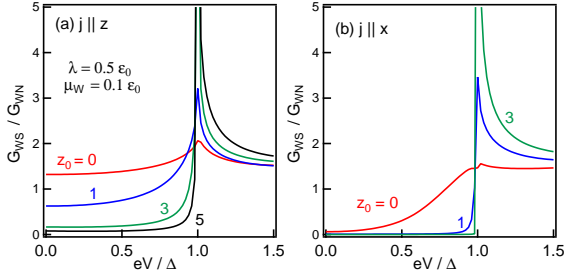}
\caption{The differential conductance of Weyl-semimetal/superconductor 
junction is plotted as a function of the bias voltage
for the current parallel to the $z$ axis in (a) and for the current parallel to the $x$ axis in (b).
Here we choose $mu_W=\mu_S$, $\mu_S=2 \epsilon_0$, $\lambda=0.5\epsilon_0$, and $\mu_W=0.1\epsilon_0$.
}
\label{fig3}
\end{figure}
The all results in (a) and (b) show the dip structure below the gap because 
the transmission probability of the junction is less than unity even at $z_0=0$ 
reflecting the difference in the band structure between the Weyl semimetal and the superconductor. 
The transmission probability $T_N$ in the normal state is 
0.76, 0.48, 0.16 and 0.076 for $z_0=0$, 1, 3 and 5, respectively in (a). 
In (b), $T_N$ is 0.71, 0.31 and 0.05 for $z_0=0$, 1 and 3, respectively. 
The gap structure becomes clearer when we decrease $T_N$ by increasing $z_0$. Such behaviors 
are well known in the conductance spectra in normal-metal/superconductor junctions. 
At the first glance, we cannot find any characteristic features of Weyl semimetals in Fig.~\ref{fig2}. 

\begin{figure}
\includegraphics[width=8cm]{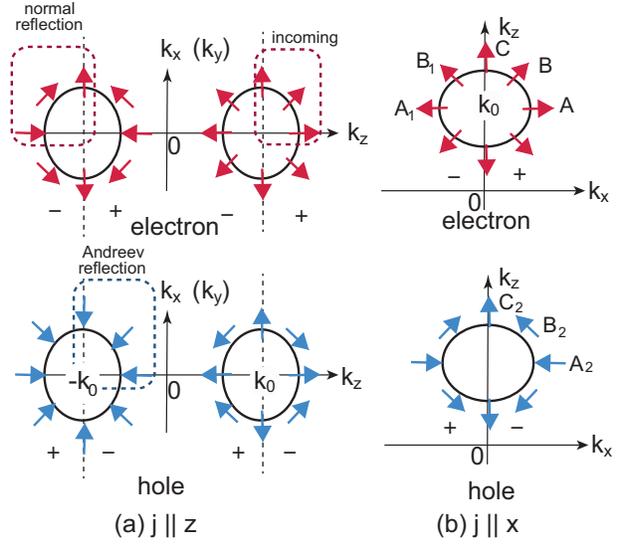}
\caption{Spin configuration on the fermi surface within $xz$ plane in spin space. 
In the junction for the current parallel to $z$,
an electron is incident from 
the two fermi surfaces near $(0,0,\pm k_0)$ having the 
opposite spin chiral texture to each other as shown in (a).
For the current parallel to $x$ in (b), an electron is reflected into the same fermi surface 
because $k_y$ and $k_z$ are conserved due to the translational symmetry in $yz$ plane.
We only show the spin configuration on the fermi surface around $(0,0, k_0)$ in (b). 
The spin is conserved in the normal reflection and it becomes opposite direction in the
Andreev reflection. 
The $+(-)$ in the bottom of figures indicates the sign of the velocity in the current direction.
}
\label{fig4}
\end{figure}

The results, however, reflect the characteristic spin configuration of Weyl semimetals.
In Fig.~\ref{fig4}, we illustrate the spin configuration on the fermi surface for 
the two types of junction. 
When the current is parallel to the $z$ direction as shown in (a), an electron 
goes into the interface from the two fermi surface near $\pm K_0$. Here we focus on an electron wave
incoming from $k_z=k_e^+>0$ channel as enclosed by the dotted rectangular. 
The spin calculated from Eq.~(\ref{spin}) are 
\begin{align}
\boldsymbol{S}(k_x,k_y,k_e^+)=\left( \frac{\lambda k_{x}}{\mu_W+E}, \frac{\lambda k_{y}}{\mu_W+E},\; 
 \frac{\sqrt{D_e^z}}{\mu_W+E} \right), 
\end{align}
for an incoming electron at $k_z=k_e^+$. 
The wave numbers in the $z$ direction for outgoing channel in the electron space 
are $k_e^-$ and $-k_e^+$. The former is in the same fermi surface as that in the incident wave,
whereas the latter belongs to the opposite fermi surface. 
In such outgoing channels, $S_x$ and $S_y$ remain unchanged from those in the incoming one. 
However, $S_z = \sqrt{D_e^z}/(\mu_W+E)$ and $-\sqrt{D_e^z}/(\mu_W+E)$ for $-k_e^+$ and   
$k_e^-$, respectively. 
In Fig.~\ref{fig4}(a), the direction of spin within $xz$ spin plane 
is illustrated by arrows on the fermi surface at $k_y=0$. 
It is possible to obtain 
the same spin configuration within $yz$ plane at $k_x=0$.
Due to the spin mismatch in $S_z$, the normal reflection to $k_e^-$ is basically suppressed.
To conserve the spin direction, therefore, an electron 
incoming at $k_e^+$ is reflected into 
$-k_e^+$ in the normal reflection.
In the outgoing channels in the hole space, the spin becomes 
\begin{align}
\boldsymbol{S}(k_x,k_y,k_h^+)= 
\left(\frac{-\lambda k_{x}}{\mu_W-E}, \;
 \frac{-\lambda k_{y}}{\mu_W-E},\; 
 \frac{\sqrt{D_h^z}}{\mu_W-E} \right)
\end{align}
at $k_z=k_h^+$, and 
\begin{align}
\boldsymbol{S}(k_x,k_y,-k_h^-)= 
\left(\frac{-\lambda k_{x}}{\mu_W-E}, \;
 \frac{-\lambda k_{y}}{\mu_W-E},\; 
 \frac{-\sqrt{D_h^z}}{\mu_W-E} \right)
\end{align}
at $k_z=-k_h^-$.
 For $E\sim \Delta \ll \mu_W$, $S_x$ and $S_y$ components in the hole space 
 are almost opposite to those in the incoming electron.  
However, $S_z$ component depends on the outgoing channels.
The Andreev reflection flips the spin of an incoming electron because 
we assume the spin-singlet superconductor. 
Thus the Andreev reflection 
is possible at $-k_h^-$ belonging to the opposite fermi surface.
An incident electron from one fermi surface is reflected in to the opposite fermi surface 
without suffering the spin mismatch. 
In the junction for the current parallel to $z$ direction, therefore,
the reflection processes are as usual as those in the normal metal. 
We note that there is no spin mismatch in the $S_y$ component. 

The situation in the junction for the current parallel to the $x$ is more complicated.
The spin of an incoming electron at $k_x=k_e^+>0$ channel is 
\begin{align}
\boldsymbol{S}(k_e^+,k_y,k_z)= \frac{1}{\mu_W+E}
\left( \lambda k_e^+, \;
\lambda k_{y},\; 
-\frac{\tilde{\lambda} \lambda k_0}{2} + \sqrt{D_e^x}\right).
\end{align}
On the other hand, they are 
\begin{align}
\boldsymbol{S}(-k_e^+,k_y,k_z)= \frac{1}{\mu_W+E}
\left(-\lambda k_e^+, \;
 \lambda k_{y},\; 
 -\frac{\tilde{\lambda} \lambda k_0}{2} + \sqrt{D_e^x}\right),
\end{align}
in the normal reflection at $k_x= -k_e^+$ in the electron space, and 
\begin{align}
\boldsymbol{S}(k_h^+,k_y,k_z)= \frac{1}{\mu_W-E}
\left( -\lambda k_h^+, \;
 -\lambda k_{y},\; 
 -\frac{\tilde{\lambda} \lambda k_0}{2} + \sqrt{D_h^x}\right),
\end{align}
in the Andreev reflection at $k_x= k_h^+>0$ in the hole space.
Although the $S_y$ component always satisfies the spin selection rule,
$S_x$ and $S_z$ components violate the selection rule depending on 
the incident angle. 
In Fig.~\ref{fig4}(b), the spin configuration within $xz$ spin plane 
is illustrated by arrows on the fermi surface. 
We only show the spin structure on the fermi surface around $k_z=k_0$
because the wave number in the $yz$ plane is conserved in the reflection process.
When the incident electron has a wave number as indicated by $A$ 
in Fig.~\ref{fig4}(b),
the normal reflection to $A_1$ is suppressed due to the spin mismatch 
in $S_x$ component but the Andreev reflection to $A_2$ is possible. 
On the other hand, when the incident electron has a wave number as indicated by $C$,
the normal reflection to $C$ itself is possible but the Andreev reflection 
to $C_2$ is suppressed because of the spin mismatch in $S_z$ component. 
 At the intermediate incident wave number as indicated by $B$, 
both the normal and the Andreev reflections are allowed.
Therefore the reflection property depends strongly on the 
incoming wave number. The results in Fig.~\ref{fig3}(b) indicate 
that the Andreev reflection probability is small near the zero-bias 
because of the spin mismatch in the reflection process.
The conductance spectra, however, are expected be sensitive to 
spin active potential at the interface.

\begin{figure}
\includegraphics[width=9cm]{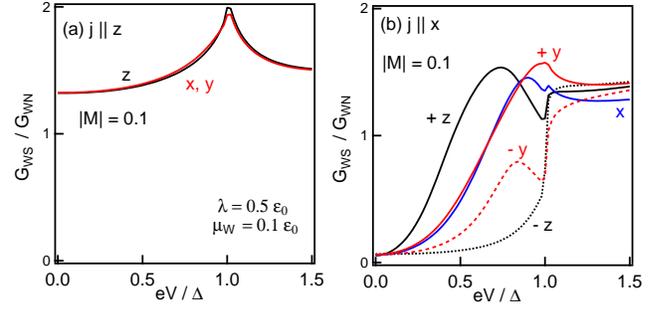}
\caption{The differential conductance 
is plotted as a function of the bias voltage
for the current parallel to the $z$ axis in (a) and for the current parallel to the $x$ axis in (b).
We introduce the magnetic barrier at $|\boldsymbol{M}|=0.1$.
Here we choose $z_0=0$, $\mu_W=0.1\epsilon_0$ and $\lambda=0.5\epsilon_0$.
}
\label{fig5}
\end{figure}

In Fig.~\ref{fig5}, we show the conductance spectra in the presence of the spin dependent
potential at the interface, where $z_0=0$ and $|\boldsymbol{M}|=0.1$. 
In (a), results for the current in the $z$ direction is plotted for three directions of 
$\boldsymbol{M}$. The conductance spectra are almost unchanged from the results 
in Fig.~\ref{fig3}(a) with $z_0=0$ because there is no spin mismatch in the reflection processes.
However, the conductance spectra for the current parallel to $x$ depends sensitively on the direction of
the magnetic moment $\boldsymbol{M}$.  
For example, $+z$ and $-z$ mean the magnetic moment points $+z$ and $-z$ direction in spin space, respectively.
The conductance for $+x$ and that for $-x$ are identical to each other.
The magnetic moment at the interface drastically modifies subgap spectra for the current 
parallel to $x$ direction because it relaxes the spin mismatch in both 
the normal and the Andreev reflection processes.

\begin{figure}
\includegraphics[width=9cm]{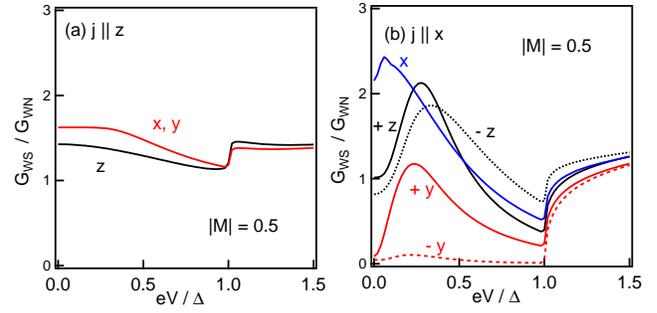}
\caption{The differential conductance at $|\boldsymbol{z}|=0.5$.
(a): the current parallel to the $z$ axis. (b): the current parallel to the $x$ axis. 
}
\label{fig6}
\end{figure}

The conductance spectra also depends on the amplitude of the magnetic moment.
In Fig.~\ref{fig6}, we show the conductance spectra for $|\boldsymbol{M}|=0.5$.
The peak at $eV=\Delta$ is suppressed by the magnetic moment for the current 
parallel to $z$ direction as shown in (a). The subgap spectra 
are totally smooth function of $eV$. In (b), on the other hand, the results show
a rich variety of the subgap spectra depending on the directions of magnetic moment.
In particular, the amplitudes of the conductance around the zero-bias for for $\pm z$ and $x$ 
are larger than those at $\boldsymbol{M}=0$ in Fig.~\ref{fig3}(b). 
This means the large amplitude of the Andreev reflection probability.

\section{Discussion}
The Andreev reflection means the penetration of a Cooper pair into the Weyl semimetal.
When the current is parallel to $z$ direction, the results in Fig.~\ref{fig3}(a) are 
qualitatively the same as those in normal-metal/superconductor junction as we discussed 
in Sec.~3. Thus a spin-singlet $s$-wave Cooper pair would be dominant in the semimetal.
When the current is parallel to $x$, on the other hand, the spin-flip scattering assists 
the Andreev reflection as shown in Figs.~\ref{fig5}(b) and \ref{fig6}(b).
Thus the spin-triplet Cooper pairs are expected in the Weyl semimetal.
In addition to this, the orbital part of a Cooper pair is also modified by the reflection.
The junction interface can mix the even-parity and odd-parity components because it 
breaks the translational invariance. Thus the odd-parity spin-triplet component 
is expected as well as the even-parity spin-singlet one. Moreover 
a Cooper pair with odd-frequency symmetry might stay in the 
Weyl semimetal in the dirty limit~\cite{bergeret,ya07}.
To resolve the pairing symmetry, we need calculate the anomalous Green function 
and analyze it. This would be an interesting issue in the future.

\section{Conclusion}
We have theoretically studied the differential conductance in the junction of Weyl-semimetal and
metallic superconductor. The Weyl semimetals have the two Weyl points in the Brillouin zone at 
$\pm\boldsymbol{K}_0$ with $\boldsymbol{K}_0 =(0,0, k_0) $. Therefore, it is possible to consider two 
different configurations of the 
junction: the current parallel to $z$ axis and the current parallel to $x$ one. 
The characteristic features of the conductance spectra for the current parallel to $z$ are 
essentially the same as those in the usual normal-metal/superconductor junctions. Namely, 
the conductance spectra becomes the bulk density of states in the superconductor 
when the normal transmission probability of the junction is low. 
In addition, the conductance spectra are insensitive 
to the weak magnetic moment at the junction interface. In this case, the chiral spin structure 
on the fermi surface does not affect the reflection process at the interface.
In the case of the junction for the current parallel to the $x$ axis, on the other hand,
the chiral spin structure on the fermi surface 
suppress the Andreev reflection depending on the incident angles of a quasiparticle. 
%The normal reflection is suppressed at some incident angle and  
%the Andreev reflection is suppressed at another incident angle. 
This feature is explained by the spin mismatch between the incoming wave and the outgoing ones.
The conductance spectra depends sensitively on the direction and the amplitudes of the 
magnetic moment at the interface because the spin flip scatterings relax the spin mismatch.
The topological bound states appear on the $xy$ surface of the Weyl semimetal. 
They, however, do not affect the low energy transport in the junctions.

\begin{acknowledgment}
%\acknowledgment

The authors are grateful to Y.~Tanaka for useful discussion.
This work was supported by the "Topological Quantum Phenomena" (No. 22103002) Grant-in Aid for 
Scientific Research on Innovative Areas from the Ministry of Education, 
Culture, Sports, Science and Technology (MEXT) of Japan.
\end{acknowledgment}

%\appendix
%\section{}

\end{document}